ADVANCING EARTH AND SPACE SCIENCE

# JGR Space Physics

**Key Points:**
- AMPERE observations reveal two classes of substorm: high- and low-latitude onsets which are weak and intense, respectively
- Intense substorms experience convection-braking in the auroral bulge; weak onsets can develop into SMC
- These results suggest a framework within which different magnetospheric modes, including sawtooth events, can be understood

**Correspondence to:**
S. E. Milan,
steve.milan@le.ac.uk

# Substorm Onset Latitude and the Steadiness of Magnetospheric Convection

S. E. Milan[1,2] , M.-T. Walach[3] , J. A. Carter[1] , H. Sangha[1] , and B. J. Anderson[4]

[1]Department of Physics and Astronomy, University of Leicester, Leicester, UK, [2]Birkeland Centre for Space Science, University of Bergen, Bergen, Norway, [3]Physics Department, Lancaster University, Lancaster, UK, [4]Johns Hopkins University Applied Physics Laboratory, Laurel, MD, USA

**Abstract** We study the role of substorms and steady magnetospheric convection (SMC) in magnetic flux transport in the magnetosphere, using observations of field-aligned currents by the Active Magnetosphere and Planetary Electrodynamics Response Experiment. We identify two classes of substorm, with onsets above and below 65° magnetic latitude, which display different nightside field-aligned current morphologies. We show that the low-latitude onsets develop a poleward-expanding auroral bulge and identify these as substorms that manifest ionospheric convection-braking in the auroral bulge region as suggested by Grocott et al. (2009, https://doi.org/10.5194/angeo-27-591-2009). We show that the high-latitude substorms, which do not experience braking, can evolve into SMC events if the interplanetary magnetic field remains southward for a prolonged period following onset. We conclude that during periods of ongoing driving, the magnetosphere displays repeated substorm activity or SMC depending on the rate of driving and the open magnetic flux content of the magnetosphere prior to onset. We speculate that sawtooth events are an extreme case of repeated onsets and that substorms triggered by northward-turnings of the interplanetary magnetic field mark the cessation of periods of SMC. Our results provide a new explanation for the differing modes of response of the terrestrial system to solar wind-magnetosphere-ionosphere coupling by invoking friction between the ionosphere and atmosphere.

## 1. Introduction

The dynamics of the magnetosphere are driven primarily by the interaction of the solar wind and embedded interplanetary magnetic field (IMF) with the terrestrial field through the process of magnetic reconnection. During periods of southward-directed IMF this excites the Dungey cycle of circulation—or convection—of the field and plasma within the magnetosphere, in which reconnection at the subsolar magnetopause creates open magnetic flux and reconnection in the magnetotail closes flux again, with a general antisunward transport of open flux and sunward return flow of closed flux (Dungey, 1961). This transport is communicated to the polar ionosphere by an electrical current system linking the magnetopause, ionosphere, and ring current (e.g., Cowley, 2000; Iijima & Potemra, 1976), resulting in an ionospheric twin-cell convection pattern (e.g., Heppner & Maynard, 1987, and references therein), which comprises antisunward plasma drift in the footprint of open field lines (known as the polar cap) and sunward plasma drift at lower latitudes.

The rate of magnetopause (or dayside) reconnection is controlled by conditions in the solar wind (e.g., Milan et al., 2012, and references therein), most importantly the orientation of the IMF. The factors controlling the occurrence and rate of magnetotail (or nightside) reconnection are less well understood, but are thought to be determined by the conditions in the plasma sheet and pressure exerted on the magnetotail by the solar wind (e.g., Boudouridis et al., 2003; Hubert, Palmroth, et al., 2006; Milan et al., 2004, 2007). Dayside and nightside reconnection can occur independently of one another, leading to changes in the open magnetic flux content of the magnetosphere, with attendant changes in the size of the ionospheric polar caps; the flux transport and convection associated with these changes is described by the expanding/contracting polar cap model (e.g., Cowley & Lockwood, 1992; Hubert, Milan, et al., 2006; Milan, 2015; Siscoe & Huang, 1985).

Often, changes in open flux content can be linked with the substorm cycle (e.g., Lockwood & Cowley, 1992; Lockwood et al., 2009; Milan et al., 2003, 2007). Substorm growth phase is the accumulation of open flux in the magnetotail lobes by dayside reconnection. The near-Earth neutral line (NENL) model of substorm onset (e.g., Baker et al., 1996; Hones, 1984) asserts that substorm expansion phase (often referred to as "substorm onset") corresponds to the formation of a reconnection X-line within the closed flux of the plasma sheet, and that this closed flux must first be pinched off (forming a plasmoid) before reconnection proceeds to close the open flux of the tail lobes. Subsequently, the recovery phase marks the antisunward motion of the NENL to form a distant neutral line. The tailward motion of the NENL is thought to be associated with the pileup of newly closed flux in the near-Earth tail, but what provokes this is unclear. The present study addresses this question.

Milan, Grocott, et al. (2009) and Grocott et al. (2009) studied the auroral intensity and the convection response of substorms with different onset latitudes, that is, substorms that accumulated different amounts of open magnetic flux prior to onset. They found that high-latitude substorms (onset above 65° magnetic latitude) tend to have a weak auroral response but lead to enhanced convection in the nightside auroral zone. On the other hand, low-latitude substorms (onset below 65° magnetic latitude) have a more intense auroral response, but counterintuitively lead to a braking of the convection flow. It was suggested by Grocott et al. (2009) that this convection-braking was produced by enhanced conductance in the more intense auroral bulge, a mechanism earlier discussed by Morelli et al. (1995).

At other times the magnetosphere can achieve similar dayside and nightside reconnection rates, leading to steady magnetospheric convection (SMC) or balanced reconnection intervals in which the open magnetic flux content remains uniform for an extended period (e.g., DeJong et al., 2008, 2018; Kissinger et al., 2012; McWilliams et al., 2008; Sergeev et al., 1996, and references therein). Sergeev et al. (1996), DeJong et al. (2008), and Kissinger et al. (2012) noted that periods of SMC are often bracketed by substorm activity, so Walach and Milan (2015) and Walach et al. (2017) examined the relationship between substorms and SMC events (SMCs) in more detail. They concluded that some SMCs are substorms that have their expansion phase prolonged by continued southward IMF, so-called "driven" substorms, whereas "classic" or "isolated" substorms are those during which the IMF turns northward shortly after onset. There is also debate as to whether northward-turnings of the IMF can trigger substorm onset (see discussion in Wild et al., 2009). It is the purpose of the current study to reexamine the link between changes in the IMF, substorms and SMCs, and the onset latitude dependence of substorm intensity.

To monitor solar wind-magnetosphere coupling, convection, and substorms, we employ measurements of the magnetosphere-ionosphere coupling currents, also known as field-aligned currents (FACs) or Birkeland currents, made by the Active Magnetosphere and Planetary Electrodynamics Response Experiment (AMPERE; Anderson et al., 2000, 2002; Coxon et al., 2018; Waters et al., 2001). The magnitude of the FACs measured by AMPERE, of which the region 1 and region 2 (R1/R2) currents identified by Iijima and Potemra (1976) are the main component, are a measure of convection strength and ionospheric conductance (Coxon et al., 2016; Milan et al., 2017), whereas the location of the R1/R2 currents is related to the open magnetic flux content of the magnetosphere (Clausen et al., 2012; Iijima & Potemra, 1978). AMPERE measurements have been used to study the large-scale magnetospheric response to solar wind driving (e.g., Anderson et al., 2014, 2018; Coxon et al., 2014a; Milan et al., 2017) and substorms (e.g., Clausen et al., 2013a, 2013b; Coxon et al., 2014b; Forsyth et al., 2018; Murphy et al., 2013).

Milan et al. (2015) applied principal component analysis (PCA) to AMPERE current maps to determine the dominant modes of response of FACs to solar wind driving. Subsequently, Milan et al. (2018) applied PCA separately to the dayside and nightside portions of the polar FAC pattern, allowing the temporal response of currents to magnetopause and magnetotail drivers to be examined. The same technique is employed in the current study.

## 2. Methodology

AMPERE (Anderson et al., 2000, 2002; Coxon et al., 2018; Waters et al., 2001, 2018) measures the FAC density in both northern and southern hemispheres, at geomagnetic latitudes above 40° with a latitudinal resolution of 1°, in 24 magnetic local time sectors, at a cadence of 2 min. The data used in this study cover the period 2010 to 2016.

The application of PCA to AMPERE observations has been described by Milan et al. (2015, 2017, 2018). An automated algorithm fits a circle to the boundary between the region 1 and 2 current ovals and the current density maps are transformed to be the same size and centered on the geomagnetic pole. The radius of the fitted circle, $\Lambda°$, measured in degrees of colatitude, is later used as a proxy for the size of the polar cap.

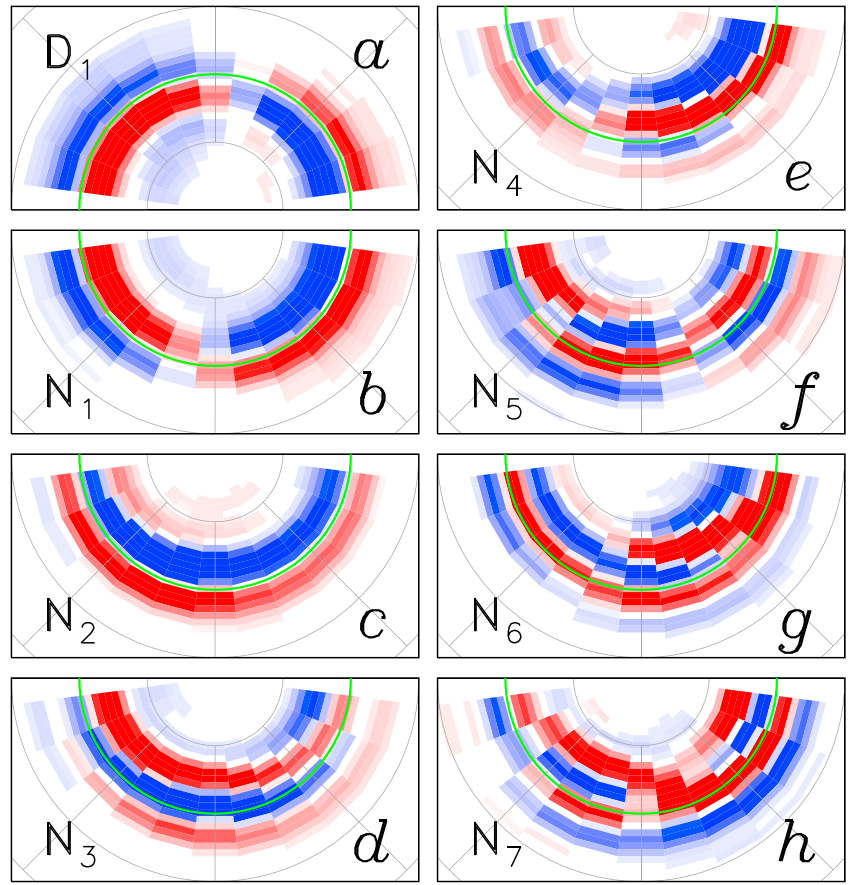

**Figure 1.** (a) The first dayside ($D_1$) and (b-h) first seven nightside ($N_i$, $i = 1, \ldots, 7$) eigenFACs derived from Active Magnetosphere and Planetary Electrodynamics Response Experiment field-aligned current distributions. Each panel is presented in a magnetic local time and magnetic latitude frame, where magnetic latitudes are scaled such that the boundary between R1 and R2 FACs occurs at 70° (green semicircle). Blue and red colors correspond to positive and negative values.

Each current map is then described by two vectors $\mathbf{J}^D$ and $\mathbf{J}^N$, being the dayside and nightside portions respectively, each of $M = 440$ elements (40 colatitudinal bins and 11 magnetic local time sectors centered on noon and midnight). PCA is then performed separately on the dayside and nightside currents, producing two sets of basis vectors $\mathbf{D}_i$ and $\mathbf{N}_i$, $i = 1, 2, 3, \ldots, M$, which are the eigenvectors of the covariance matrices of $\mathbf{J}^D$ and $\mathbf{J}^N$, respectively. These basis vectors, which we term eigenFACs, are those that most efficiently describe variations in the observations. There are as many dayside and nightside eigenFACs as there are elements in the original vectors, but only the first few are significant. The first dayside and the seven most important nightside eigenFACs are presented in Figure 1. For the time being, we note that $\mathbf{D}_1$ and $\mathbf{N}_1$ resemble the dayside and nightside portions of the R1 and R2 current systems.

Each of the original vectors $\mathbf{J}^D$ or $\mathbf{J}^N$ can be reconstructed as a linear combination of the eigenFACs:

$$\mathbf{J}^D = \sum_{i=1}^{m} \alpha_i \mathbf{D}_i, \qquad \mathbf{J}^N = \sum_{i=1}^{m} \beta_i \mathbf{N}_i, \tag{1}$$

where $\alpha_i$ and $\beta_i$ are coefficients which can be determined by finding the projection of $\mathbf{D}_i$ and $\mathbf{N}_i$ on $\mathbf{J}^D$ and $\mathbf{J}^N$:

$$\alpha_i = \mathbf{J}^D \cdot \mathbf{D}_i, \qquad \beta_i = \mathbf{J}^N \cdot \mathbf{N}_i. \tag{2}$$

For the reconstruction to be exact, all eigenFACs must be included in the summations, that is, $m = M$, though in practice reasonable fidelity can be achieved with $m \ll M$. The coefficients $\alpha_i$ and $\beta_i$ then provide a means of quantifying a complex data set using a handful of numbers, a technique known as dimensionality reduction. In this study we use $m = 1$ on the dayside and $m = 7$ on the nightside. The significance of each eigenFAC (the amount of variance in the original data that it represents) is given by the ratio of its corresponding eigenvalue to the sum of the eigenvalues of all the eigenFACs, indicated in Figure 1 of Milan et al. (2018). As shown in that figure, eigenFACs $N_1$ to $N_7$ capture ~80% of the variance in the nightside FAC patterns, with the contribution of the first 10 eigenFACs to the variance being 45.8%, 9.2%, 6.3%, 5.2%, 4.1%, 3.3%, 3.0%, 2.7%, 2.5%, and 2.2%. There is no clear change in significance between $N_7$ and $N_8$, and the choice of this cutoff in our analysis is somewhat arbitrary; however, we find that no new information for the present study is contributed by including $N_8$ and beyond. We expect $\alpha_1$ and $\beta_1$ to quantify the strengths of the dayside and nightside portions of the R1/R2 current system in each AMPERE FAC map. As we will demonstrate, $\beta_3$ to $\beta_7$ are related to substorm phenomena.

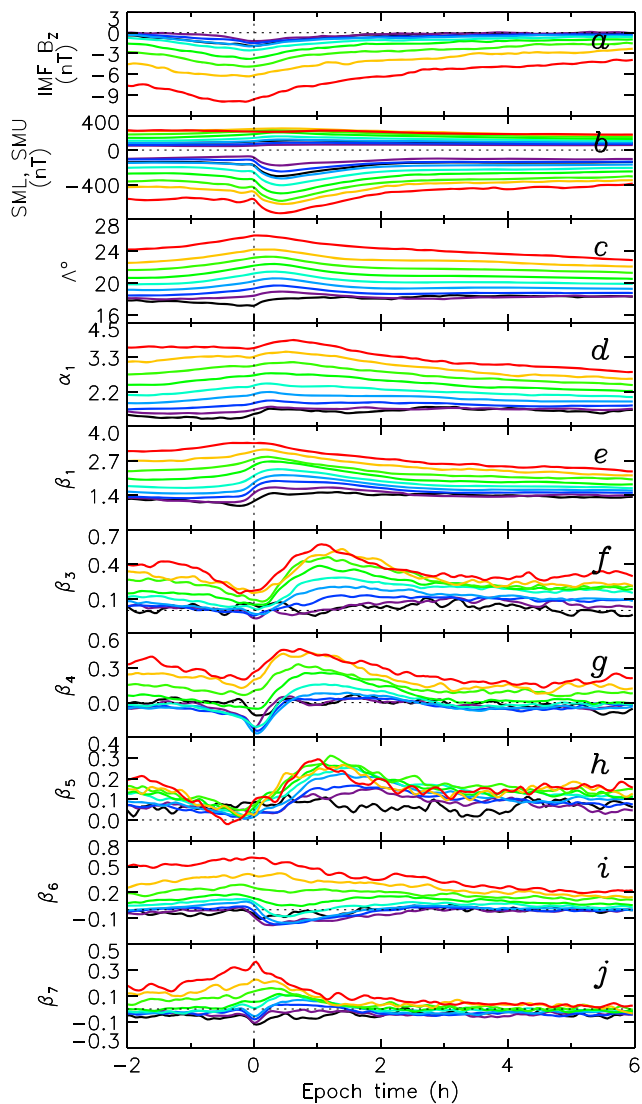

**Figure 2.** Superposed epoch analysis of substorms from 2 hr prior to 6 hr after onset. (a) IMF $B_Z$, (b) SMU and SML electrojet indices, (c) radius of the Active Magnetosphere and Planetary Electrodynamics Response Experiment current pattern, $\Lambda$, and (d to j) coefficients associated with the eigenFACs presented in Figure 1 (except $N_2$). The substorms are categorized by the value of $\Lambda$ at onset ($t = 0$) or $\Lambda_0$, in 1° steps from 18° (black trace) to 26° (red trace); the number of substorms in each category is 976, 510, 1282, 1681, 1527, 1108, 789, 501, and 522, respectively. IMF = interplanetary magnetic field.

Supporting data are provided by the 1-min OMNI data set (King & Papitashvili, 2005); the SuperMAG geomagnetic index data set, including SML, SMU, and SMR, equivalent to AL, AU, and SYM-H (Gjerloev, 2012; Newell & Gjerloev, 2011, 2012); and the substorm onset list derived from SuperMAG observations (Newell & Gjerloev, 2011).

## 3. Observations and Discussion

This paper focusses on three aspects of solar wind-magnetosphere coupling, substorms, and substorm-related FACs: How does the FAC response vary with substorm onset latitude? What is the relationship between substorms and SMC? And what do the FAC systems tell us about magnetosphere-ionosphere coupling during substorms? We investigate these three themes in turn.

Figure 2 presents a superposed epoch analysis of substorms, from 2 hr before to 6 hr after substorm onset. Panels (a) to (c) show IMF $B_Z$, the electrojet indices SML and SMU, and the radius of the boundary between the R1/R2 current ovals, $\Lambda$°, a proxy for polar cap size. Panels (d) to (j) show the coefficients $\alpha_1$ and $\beta_1$ to $\beta_7$, neglecting $\beta_2$. As described by Milan et al. (2018), $\beta_2$ quantifies IMF $B_Y$ asymmetries in the nightside FACs and is not of interest to the present study. The substorms are categorized by the value of $\Lambda$ at the time of onset, $\Lambda(t = 0)$ or $\Lambda_0$, and the corresponding traces color-coded from $\Lambda_0 = 18°$ (black) to $\Lambda_0 = 26°$ (red) in steps of 1°. In total, 8,896 substorms are included in the analysis. For clarity, the traces do not show the standard error on the mean, though these are in general low due to the relatively large number of substorms in each category.

On average, IMF $B_Z$ is close to zero or negative throughout the period of analysis. This is because substorms tend to occur during periods of southward IMF which lead to magnetopause reconnection and substorm growth phase. More negative values of $B_Z$ are associated with lower-latitude (higher-$\Lambda_0$) onsets. That is, substorms tend to occur on an expanded auroral oval, corresponding to high polar cap flux, when $B_Z$ is strongly southward. $B_Z$ becomes more negative as onset is approached, associated with substorm growth phase, and less negative afterward, as there is no longer a requirement for continued creation of open flux after the substorm has commenced. The SMU and SML indices (the SuperMAG equivalents of the AU and AL electrojet indices) show substorm growth, expansion, and recovery phase signatures, as expected, though the magnitude of the variations are larger for high-$\Lambda_0$ onsets. In all categories except the lowest-$\Lambda_0$ substorms, $\Lambda$ increases prior to onset, a signature of growth phase, and decreases thereafter. The beginning of the contraction of the polar cap can be delayed by almost an hour after onset for the low-$\Lambda_0$ substorms. Panels (d) and (e) show the strength of the dayside and nightside R1/R2 currents, as quantified by $\alpha_1$ and $\beta_1$, which quantify the rate of convection on the dayside and nightside (e.g., Clausen et al., 2013a; Milan, 2013). The magnitude of the R1/R2 FACs is well ordered by $\Lambda_0$, indicating that magnetospheric convection is enhanced for low-latitude onsets. The dayside R1/R2 tends to grow during the growth phase, and then steps up following onset, before decaying after a few tens of minutes. The nightside R1/R2 FACs remain roughly constant during the growth phase but again increase around the time of onset. These results are broadly consistent with previous studies of the latitude dependence of substorms (e.g., Clausen et al., 2013b; Coxon et al., 2014b; Milan, Grocott, et al., 2009).

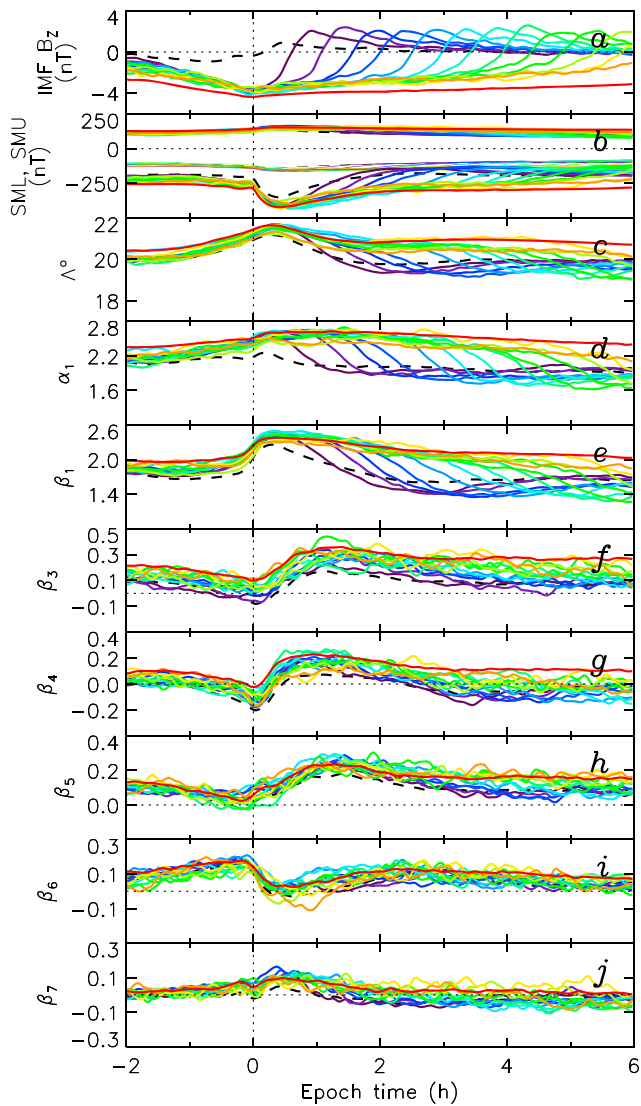

**Figure 3.** Superposed epoch analysis of substorms in the same format as Figure 2. Substorms are categorized by the length of time that the IMF remains southward after onset: over 30, 60, 90, …, 360 min (blue to red traces). The number of substorms in each category is 699, 495, 472, 344, 337, 290, 253, 229, 234, 184, 222, 151, 189, and 2,786, respectively. Substorms which do not fit the selection criteria (2,409) are shown as a dashed line. IMF = interplanetary magnetic field.

Panels (f) to (j) show the nightside response of the FACs as quantified by $\beta_3$ to $\beta_7$. We defer a discussion of what the eigenFACs $N_3$ to $N_7$ actually signify until later in the paper. For the time-being we note that all five parameters show substorm-related variations, that is, their behavior shows marked changes before, during, and after onset. Their variations are also ordered by $\Lambda_0$: specifically, there appear to be two classes of behavior displayed by low- and high-$\Lambda_0$ substorms. For instance, for $\Lambda_0 < 21°$ (black to cyan traces) $\beta_4$ decreases from 0 in the 30 mins before onset and increases back to 0 in the 30 mins after onset; for $\Lambda_0 > 21°$ (green to red traces) $\beta_4$ shows little variation prior to onset, but increases for an hour or so after onset. Similar, clear differences between these two latitudinal classes can be seen in the variations of $\beta_6$ and $\beta_7$.

Grocott et al. (2009) also identified two classes of substorm: those that experience convection-braking (onsets below 65° magnetic latitude) and those that do not (onsets above 65°). Our value of $\Lambda_0$ of 21° is consistent with 65° at midnight, as the auroral oval is on average displaced antisunward from the geomagnetic pole by 4°. Their interpretation was that enhanced conductance, associated with enhanced auroral luminosity for low-latitude onsets (Milan, Grocott, et al., 2009), leads to frictional coupling between the ionosphere and atmosphere such that the convection is arrested. Following on from the results of Grocott et al. (2009), in the remainder of this study we assume that our low-latitude category of onsets experience convection-braking, whereas our high-latitude onsets do not. We will go on to demonstrate that high-latitude onsets can evolve into periods of SMC, but that low-latitude onsets cannot. (We note that DeJong et al., 2018, presented a counter-example to this hypothesis, in which a case study of the conductance during an isolated substorm and an SMC event indicated higher conductance during the latter.)

We next investigate the nature of convection associated with substorms during which IMF $B_z$ remains southward for different lengths of time following onset. In general $B_z < 0$ nT during the growth phase—it is generally accepted that ongoing loading of open flux into the magnetosphere is a prerequisite for substorm onset, unless an external perturbation such as a solar wind pressure pulse triggers onset (e.g., Boudouridis et al., 2003; Hubert, Palmroth, et al., 2006; Milan et al., 2004)—but once onset has commenced, the IMF orientation can change. Figure 3 presents a superposed epoch analysis of substorms in the same format as Figure 2 (though note that the vertical scales differ between the two figures). In this analysis, substorms are categorized by the length of time that IMF $B_z$ remains negative after onset. In each category, we require that $B_z < -2$ nT for 90% of data points from 20 min before onset to 30, 60, 90, …, 360 min after onset, with traces color-coded from black to red. Substorms which do not match these criteria are indicated by a dashed line.

The resulting $B_z$ traces show the expected variation, becoming increasingly negative prior to onset, and then remaining negative for a different length of time post-onset before turning positive. The corresponding

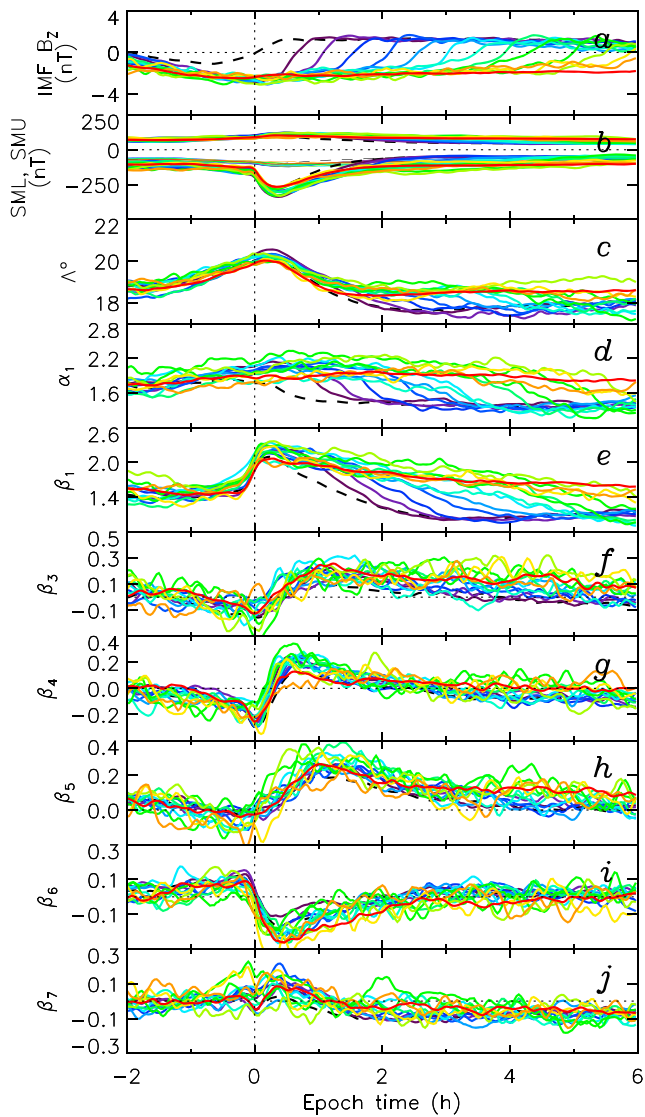

**Figure 4.** Superposed epoch analysis of substorms in the same format as Figure 2, and selected in the same way, though restricted to those substorms which are not followed by a subsequent substorm for at least 6 hr. The number of substorms in each category is 59, 180, 128, 85, 97, 58, 48, 65, 37, 35, 27, 26, 31, and 306, respectively. Substorms which do not fit the selection criteria (847) are shown as a dashed line. IMF = interplanetary magnetic field.

SMU and SML traces show the expected substorm growth, expansion, and recovery phases, except that the duration of the substorm bay in SML is prolonged by the length of time that $B_Z$ remains southward. The radius of the current ovals, $\Lambda$, increases during the growth phase prior to onset and begins to fall about 20 min after onset, but remains elevated for the duration of the $B_Z$-southward phase. Similar behaviors are seen for the dayside and nightside R1/R2 FAC magnitudes as quantified by $\alpha_1$ and $\beta_1$: increasing before onset and remaining elevated during the period of southward IMF before falling to pre–growth phase levels, that is, convection strength increases during the growth phase and is maintained while the magnetosphere continues to be driven. At this point, with regards to the variations of $\beta_3$ to $\beta_7$, we note that these are similar to each other for all categories of $B_Z$-southward duration.

On the face of it, these results would seem to support the conclusions of Walach and Milan (2015), that continuing southward IMF after substorm onset can lead to a period of SMC, which only subsides once the IMF turns northward. However, we have not considered the possibility that with continued southward IMF a series of substorms could be triggered, and that averaging over many such substorms could lead to the results presented in Figure 3. To investigate further, we repeat the superposed epoch analysis, but now limit the events to those substorms for which there is no subsequent substorm in the following 6 hr. This significantly reduces the number of events in the analysis, so we relax our $B_Z$ criterion to be that $B_Z < -1$ nT (rather than $B_Z < -2$ nT) for 90% of data points from 20 min before onset to 30, 60, 90, ..., 360 min after onset. The results are presented in Figure 4.

The $B_Z$ traces are similar to Figure 3, though $B_Z$ is not as negative as before. The SML traces for each category are similar to each other, indicating a substorm bay that lasts 90 min in each case—a single expansion phase lasting approximately 1 hr irrespective of the duration of the $B_Z$-southward phase. However, after the expansion and contraction of the polar cap associated with the onset, $\Lambda$ remains elevated for the duration of the $B_Z$-southward phase. Similarly, the dayside and nightside R1/R2 current magnitudes are also elevated for the duration of the $B_Z$-southward phase. These results do appear to confirm the conclusions of Walach and Milan (2015), that nightside reconnection can be maintained at the end of the expansion phase of a substorm, and steady convection can ensue, if the IMF remains southward.

We now compare other differences between the substorms of Figures 3 and 4. First, Figure 3 has an average value of $\Lambda_0 > 21°$ while for Figure 4, on average $\Lambda_0 < 21°$. The difference in the two average values is marginal, but does place the two sets of substorms in the high- and low-$\Lambda_0$ categories discussed in relation to Figure 2. Moreover, in Figure 4 both $\beta_4$ and $\beta_6$ become negative at the time of onset and shortly afterward, respectively, whereas this negative excursion is not so significant in Figure 3. This reinforces the link between the high- and low-$\Lambda_0$ categories and Figures 3 and 4, respectively, that is the dip in $\beta_4$ and $\beta_6$ distinguishes those substorms that do not experience convection-braking from those that do.

Figure 5 presents a schematic of the two scenarios we envisage for substorms occurring during prolonged periods of solar wind-magnetosphere coupling, with high-latitude substorm onsets on the left, panels (a)–(c), and low-latitude onsets on the right. The figure has a format similar to Figure 3 of Cowley and Lockwood (1992). Panels (a) and (d) show substorm growth phase for the two cases, followed by the expansion phase in panels (b) and (e). We suggest that substorms that can transition to periods of SMC (panel (c)) are those that do not experience convection-braking, whereas substorms that do experience braking cannot lead to a laminar convection state, but must result in a sequence of onsets if the IMF remains southward (panel (f)).

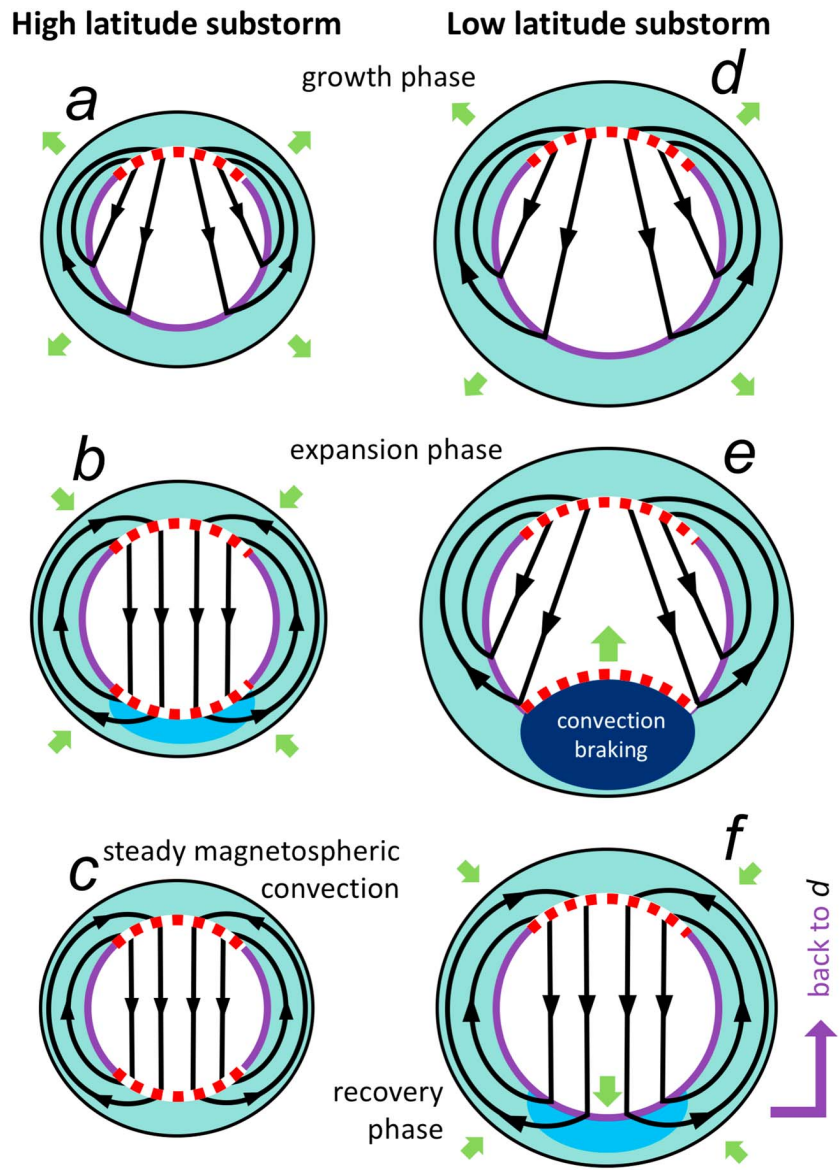

**Figure 5.** A schematic of the development of high- and low-latitude substorms, panels (a) to (c) and (d) to (f), respectively, in response to prolonged solar wind-magnetosphere coupling. Each panel has noon at the top. Black arrowed lines show convection streamlines, the purple circle is the OCB enclosing the polar cap. Red dashed lines show portions of the OCB mapping to active reconnection X-lines at the magnetopause or in the magnetotail. Blue regions indicate the location of the auroral bulge and whether it is of high (dark blue) or low (light blue) ionospheric conductance; along these portions of the OCB the ionospheric flow crosses the boundary, along other portions the flow is adiaroic. Green arrows indicate expansion or contraction of the auroral zone and polar cap, or motions of the OCB. OCB = open/closed field line boundary.

An expected consequence of convection-braking is the formation of a pronounced substorm auroral bulge following onset, with a significant poleward motion of the nightside open/closed field line boundary (OCB) as magnetotail reconnection erodes the open flux of the polar cap (panel (e)). As the bulge begins to dim and the brake is released, the substorm enters a recovery phase in which the polar cap returns to a circular shape through the redistribution of open and closed flux in the ionosphere (panel (f)), before the cycle begins again (panel (d)). Conversely, we expect that substorms with no convection-braking can maintain a roughly circular polar cap through continuous redistribution of flux, such that the substorm appears as a brightening of the nightside auroral oval rather than a poleward-growing bulge (panel (b)). There is evidence in the wideband imaging camera observations of Figure 5 of Milan, Grocott, et al. (2009) to support this suggestion that low-latitude onsets have a much more significant poleward progression of the substorm auroras than high-latitude onsets, and this is also consistent with the nightside auroral observations of an isolated substorm and an SMC event presented by DeJong et al. (2018).

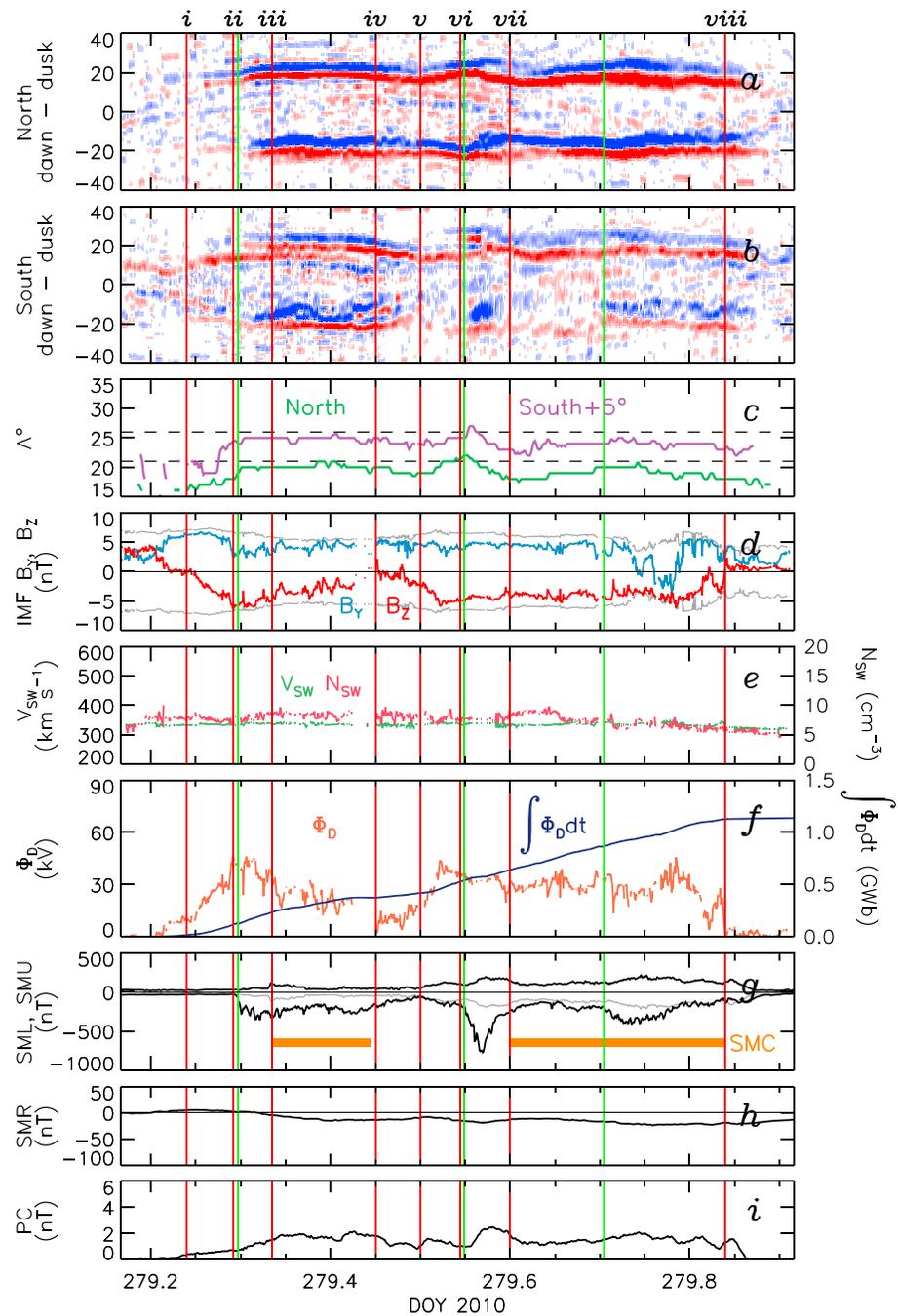

**Figure 6.** Solar wind and magnetospheric dynamics for the period 04 to 22 UT, 6 October 2010. (a) and (b) Field-aligned current density along the dawn-dusk meridian in the northern and southern hemispheres, with red and blue indicating upward and downward field-aligned current, with the color scale saturating at 0.5 $\mu$A/m$^2$. (c) Radius, $\Lambda$, of the northern and southern field-aligned current ovals, with the southern hemisphere being displaced by 5° for clarity; the horizontal dashed lines indicate $\Lambda = 21°$. (d) IMF $B_Y$ and $B_Z$. (e) Solar wind speed and density. (f) Dayside reconnection rate, $\Phi_D$, and the time integral of $\Phi_D$. (g) The electrojet indices SML and SMU, with periods of steady magnetospheric convection indicated by orange bars. (h) The ring current index SMR. (i) The PC index. Vertical green lines show SuperMAG substorm onsets, and vertical red lines are discussed in the text. IMF = interplanetary magnetic field.

Convection-braking is also expected to have ramifications for the dynamics in the magnetotail. The substorm onset marks the formation of a NENL in the closed plasma sheet (e.g., Baker et al., 1996; Hones, 1984). Once the NENL has pinched off the closed flux, open flux is closed and the polar cap contracts. If the redistribution of magnetic flux in the ionosphere required by the expanding/contracting polar cap is unimpeded, then the NENL reconnection rate can adjust to match the dayside reconnection rate and a period of SMC, that is a balanced reconnection interval, can ensue. On the other hand, if convection-braking occurs, ongoing tail reconnection will lead to the formation of a poleward-progressing auroral bulge. This will be associated with flux pileup in the near-tail, as newly closed flux cannot convect onward, and this pileup will push the NENL down-tail until reconnection ceases. Reconnection can only recommence by the formation of a new NENL within the region of newly closed field lines, signaled by a new substorm AL bay. In this manner, a sequence of onsets is required if the IMF remains directed southward. In both substorm and SMC cases, once the IMF turns northward dayside reconnection ceases but nightside reconnection continues until the tail reaches a relaxed configuration.

So far we have discussed statistical results. We now present some case examples. To aid with event selection, we developed an algorithm to identify potential periods of SMC from the SMU and SML indices, using criteria similar to McWilliams et al. (2008) and Kissinger et al. (2012). We then discarded events during which the IMF was variable or the FAC ovals showed large changes in radius (see also: Walach & Milan, 2015). Many events were found, of which some typical examples are shown in Figures 6 to 8. We first discuss the 18-hr period beginning 04 UT, 6 October 2010, presented in Figure 6.

Panels (a) and (b) show AMPERE FAC densities along the dawn-dusk meridians of the northern and southern hemispheres, in which the upward/downward FAC pairs (the R1/R2 FACs) can be seen at dawn and dusk. The radii of the FAC ovals, $\Lambda$, are shown in panel (c), followed by IMF $B_Y$ and $B_Z$ and solar wind speed and density in panels (d) and (e). Panel (f) shows the dayside reconnection rate, $\Phi_D$, estimated from the solar wind observations using equation (15) of Milan et al. (2012), and the time integral of $\Phi_D$. This integral shows the amount of open flux that would accumulate in the polar caps if no nightside reconnection took place. Typically the polar caps contain 0.5 GWb of open flux, rising to ~1 GWb during extreme conditions (Milan et al., 2007). Below this are: panel (g) the SML and SMU indices, panel (h) the SMR ring current index, and panel (i) the PC index which measures convection strength in the polar regions. Vertical green lines indicate substorm onset identified by SuperMAG. Vertical red lines, labeled $i$, $ii$, and so forth, indicate times which will be discussed below; if a red line corresponds to a substorm onset, it has been displaced slightly for clarity.

Two events occurred during this time period. Between $i$ and $viii$, the IMF was predominantly directed southward (panel (d)), the R1/R2 FACs were enhanced (panels (a) and (b)), and the PC index was elevated (panel (i)). Following the southward turning at $i$, dayside reconnection was elevated leading to expansions of the polar caps (panel (c)). At $ii$, SuperMAG identified a substorm onset. Thereafter, the IMF remained southward until $iv$, and between $iii$ and $iv$ SMC ensued, indicated by the horizontal orange bar, during which SMU and SML were approximately constant (panel (g)), the FAC radius remained uniform, and PC indicated steady convection. By $v$ the IMF had turned southward again, growth phase signatures were seen in SMU/SML and $\Lambda$, followed by a substorm onset at $vi$. Associated with the substorm bay, the polar caps initially contracted, but by $vii$ they stabilized and varied only gradually during a second period of SMC, again accompanied by steady PC. At $viii$ the IMF rotated so that it was no longer southward, and the SMC petered out. We consider both these periods of SMC to be examples of the driven-substorm SMC described by Walach and Milan (2015). We note that $\Lambda$ remained below 21° throughout almost the entire period.

To summarize these two events, we identify the following intervals as follows: ($i - ii$) growth phase, ($ii - iii$) expansion phase, ($iii - iv$) SMC, ($iv - v$) recovery phase, ($v - vi$) growth phase, ($vi - vii$) expansion phase, ($vii - viii$) SMC, ($viii-$) recovery phase. In both of these cases, $\int \Phi_D dt \sim 0.1$ GWb of open flux accumulated during the growth phase of the substorm and ~0.1 GWb during the expansion phase. During the two periods of SMC, ~0.15 and ~0.5 GWb of open flux were open and closed, that is, transported through the system, in the latter case equivalent to refreshing a typical polar cap.

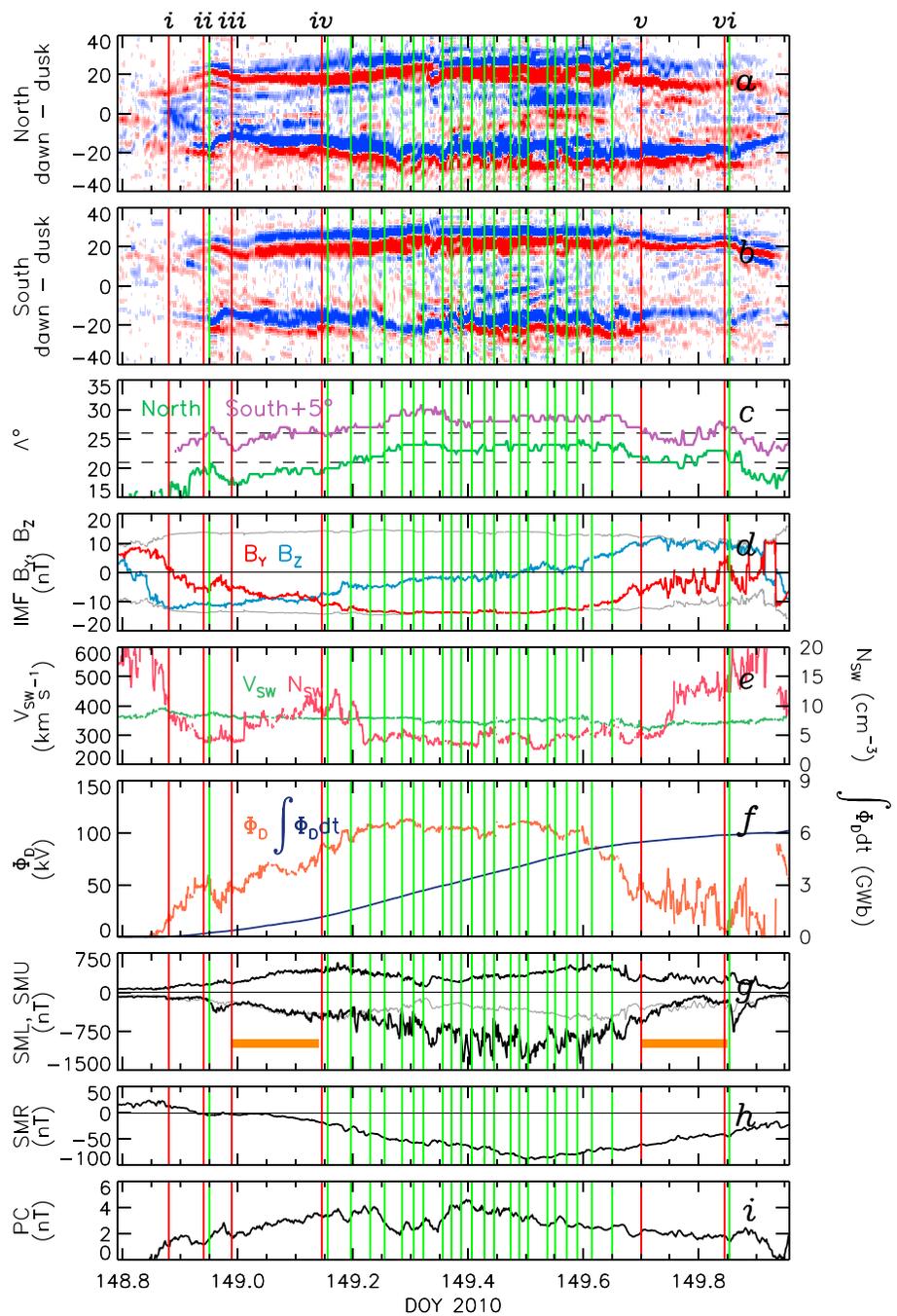

**Figure 7.** Solar wind and magnetospheric dynamics for the period 19 UT, 28 May 2010, to 23 UT on the following day, in the same format as Figure 6.

Figure 7 presents the 28-hr period beginning 19 UT, 28 May 2010. At *i* the IMF turned southward and remained so for almost 23 hr. Following *i* the polar caps expanded (Λ), before a substorm onset was detected at *ii*, which by *iii* developed into a period of SMC. IMF $B_Z$ became increasingly negative such that the dayside reconnection rate increased and exceeded the nightside rate resulting in gradually expanding polar caps (Λ) and stronger convection (SML/SMU and PC). Around *iv*, Λ grew beyond 21° and thereafter multiple substorm onset or substorm intensification signatures were identified by SuperMAG. By *v*, the IMF was no longer so strongly southward, the polar caps had contracted, and SMC resumed, until the IMF turned northward at *vi*. A substorm occurred at this time and the polar caps rapidly contracted. This event shows that SMC can occur when the polar caps are contracted, but if the radii grow too large, repeated substorm activity results. We also note that this period of elevated polar cap size is associated with an enhanced ring current (SMR, similar to SYM-H), as suggested by Milan, Hutchinson, et al. (2009). $\int \Phi_D dt \sim 0.2$ GWb during both the growth and expansion phases of the initial substorm, and $\sim 0.8$ and $\sim 0.4$ GWb during the two phases of SMC. Between *iv* and *v*, $\int \Phi_D dt \sim 4.5$ GWb, or approximately 0.2 GWb for each intensification. It is debatable if each substorm onset identified by SuperMAG in the interval *iv* to *v* is a true substorm or rather a substorm intensification; however, this clearly is not a period of SMC, and SML indicates intense fluctuations in nightside precipitation which would be expected to give rise to convection-braking.

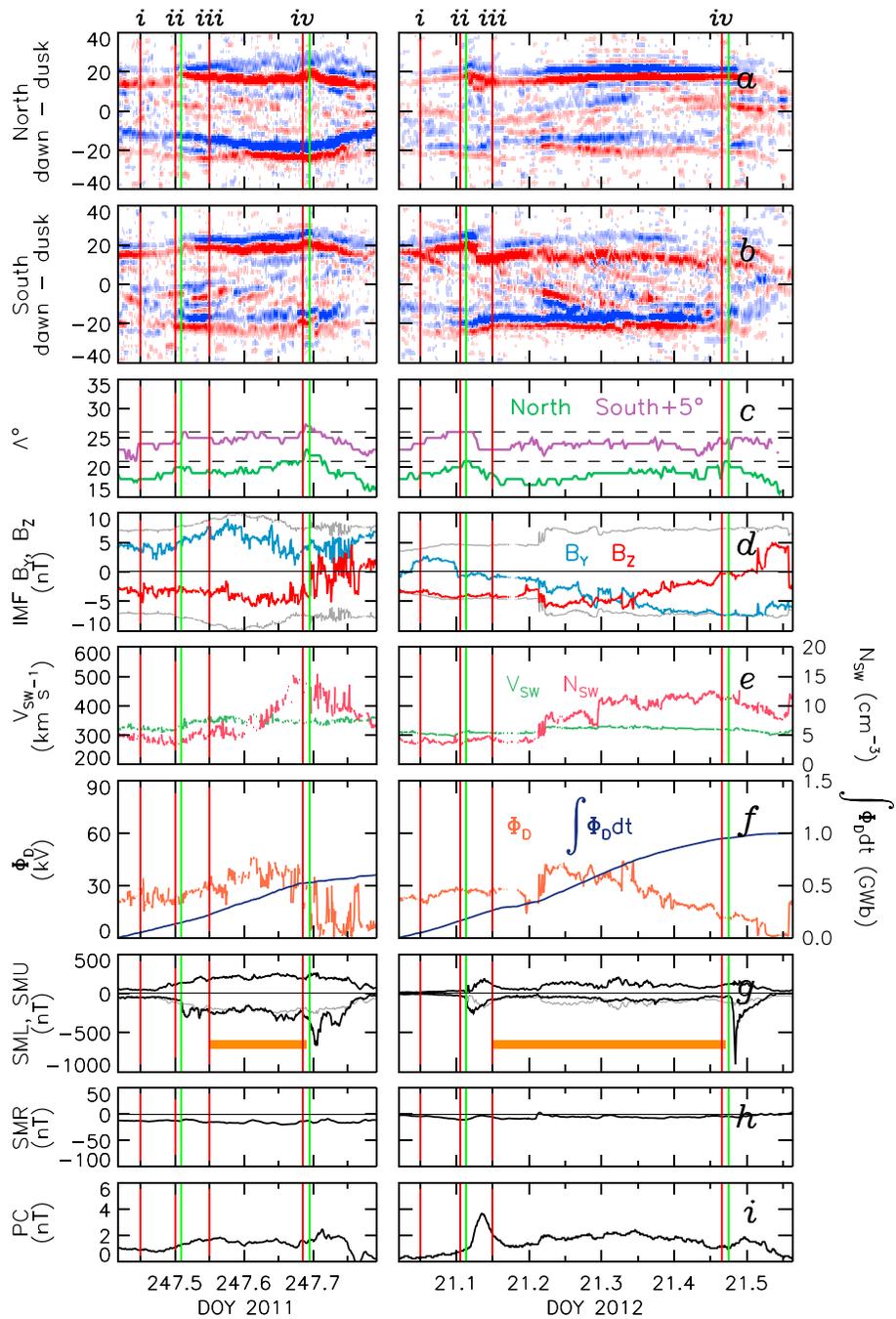

**Figure 8.** Solar wind and magnetospheric dynamics or the period 10 to 19 UT, 4 September 2011, and 00:30 to 13:30 UT, 21 January 2012, in the same format as Figure 6.

Finally, Figure 8 presents two similar events: the 9-hr period after 10 UT, 4 September 2011, and the 13-hr period after 00:30 UT, 21 January 2012. In both examples, during ongoing southward IMF growth phase signatures were observed between *i* and *ii*, followed by substorm onset at *ii*, transitioning into SMC at *iii*, and then ending as the IMF turned northward at *iv*. In both cases a substorm onset was observed at *iv*, which lead to rapid contractions of the polar caps. Several studies have indicated that periods of SMC often end with a substorm (e.g., Sergeev et al., 1996), and these examples (and arguably that in Figure 7) conform to this. In none of these cases is there a clear solar wind cause for the triggering of a substorm, except a reduction in the dayside reconnection rate. We suggest that if the tail is in a stressed state at the end of a period of SMC, a substorm can be triggered to return it to a relaxed state. In all three cases the solar wind density is relatively high, exceeding 10 cm$^{-3}$, which may exacerbate this stressed state. Interestingly, such cases may inform the debate regarding the existence of substorms triggered by northward turnings of the IMF (e.g, Wild et al., 2009, and references therein). In both these cases, $\int \Phi_D dt \sim 0.1$ GWb during the growth and expansion phases of the initial substorms (i.e., ~0.2 GWb associated with each substorm), and ~0.3 and ~0.7 GWb during the two phases of SMC.

We now turn to our final question, regarding the nature of the FAC patterns associated with substorm onset, and especially the difference between the high- and low-$\Lambda_0$ onsets. Remembering that a general FAC pattern can be reproduced as a linear combination of the eigenFACs, we briefly describe the FAC morphology associated with $\mathbf{N}_i$, $i = 1, \ldots, 7$, and their contribution to the summation in equation (1).

$\beta_1 \mathbf{N}_1$: The nightside portion of the large-scale R1/R2 current system. $\beta_1$ is always found to be positive, as this corresponds to the expected polarity of the R1/R2 FACs. This eigenFAC is roughly symmetric about the midnight meridian, but we note that the upward FACs link up across midnight, the average configuration associated with the Harang discontinuity (e.g., Iijima & Potemra, 1978).

$\beta_2 \mathbf{N}_2$: As discussed by Milan et al. (2018), this eigenFAC controls the local time at which the polarities of the R1/R2 currents reverse, which we can identify with the location of the nightside convection throat. For $\beta_2 > 0$ and $\beta_2 < 0$ the convection throat moves pre- and post-midnight, respectively. Milan et al. (2018) showed that the value of $\beta_2$ is related to the $B_Y$ component of the IMF.

$\beta_3 \mathbf{N}_3$: This eigenFAC controls how the pre- and post-midnight portions of the R1/R2 FACs link up across midnight. If $\beta_3 < 0$ then the strength of the upward current in the Harang discontinuity region is enhanced. If $\beta_3 > 0$ then upward current is diminished or downward current intrudes into this region. We observe that $\beta_3$ tends to be positive for high-$\Lambda_0$ substorms, which is consistent with Figure 15 of Iijima and Potemra (1978).

$\beta_4 \mathbf{N}_4$: When $\beta_4 > 0$ this eigenFAC leads to a strengthening and poleward motion of the R1/R2 FACs, especially at midnight and in the post-midnight sector; $\beta_4 < 0$ leads to a thinning of these currents. That $\beta_4 > 0$ and $\beta_4 < 0$ for high- and low-$\Lambda_0$ substorms (e.g., Figure 2) is consistent with our assertion that the auroral bulge is enhanced and protrudes poleward during high-$\Lambda_0$ substorms.

$\beta_5 \mathbf{N}_5$: In the midnight sector this eigenFAC is morphologically similar to $\mathbf{N}_3$ (though of opposite polarity), so we expect it to play a role in modulating the Harang discontinuity currents.

$\beta_6 \mathbf{N}_6$: When $\beta_6 > 0$ this eigenFAC contributes upward FAC at high latitudes, especially in the midnight and post-midnight regions, and in this respect is similar to $\mathbf{N}_4$.

$\beta_7 \mathbf{N}_7$: When $\beta_7 > 0$ this eigenFAC contributes upward FAC at high latitudes, across both pre- and post-midnight regions.

In summary, those eigenFACs which tend to be enhanced during high-$\Lambda_0$ substorms contribute FAC at high latitudes in the pre-, post-, and midnight regions, consistent with our expectations that these substorms have an enhanced auroral bulge that will lead to convection-braking; indeed, the poleward growth of the bulge is a consequence of this convection-braking, requiring a poleward motion of the ionospheric projection of the nightside reconnection X-line as open magnetic flux is eroded.

## 4. Conclusions

We have examined the FAC strength and morphology during substorms, using observations from AMPERE, focusing on two main questions: What influence does onset latitude have on the FAC response? and What is the relationship between substorm onset, prolonged IMF $B_Z$-southward conditions, and SMC?

Milan, Grocott, et al. (2009) demonstrated that substorms occurring at low latitudes (high-$\Lambda_0$ substorms in the terminology of this paper) are more intense than high-latitude substorms, and Grocott et al. (2009) demonstrated that these experience convection-braking, possibly associated with the high conductance of the bright auroral bulge (e.g., Morelli et al., 1995). Walach and Milan (2015) showed that a significant number of SMC events are initially substorms, but substorms for which the IMF remains southward for a prolonged period after onset. Our results support both of these conclusions, but we go further to suggest that those substorms which can evolve into SMC are those that occur at high latitudes and do not experience convection-braking, as illustrated in Figures 5a–5c. In this case, once a substorm commences, associated with the onset of magnetic reconnection in the tail at a NENL, that reconnection can persist if new open flux continues to be supplied by dayside reconnection. Typical substorm signatures, such as the SML bay and substorm-associated FAC morphologies, last 60 to 90 min after onset, but these die away even if NENL reconnection continues thereafter. We suggest that substorms which experience braking and associated flux pile up in the near-tail pushing the NENL down-tail, cannot segue into a laminar convection state, and instead a staccato sequence of substorm onsets results, each with a recovery phase that represents the release of the brake, as illustrated in Figures 5d–5f. Substorms that experience braking will be those that develop poleward-growing auroral bulges, whereas high-latitude, nonbraking substorms will have less-pronounced bulges, maintaining a circular polar cap through frictionless redistribution of magnetic flux.

In the examples presented, between 0.2 and 0.4 GWb of open flux transport were associated with the growth and expansion phases of each precursor substorm, with between 0.15 and 0.8 GWb during the following period of SMC. This latter value depends on how long the IMF remains southward following the initial onset, that is, the duration of the SMC. When the polar caps grew sufficiently large that a sequence of substorms or intensifications was triggered, each effected 0.2 GWb of flux transport. We note that periods of SMC can lead to a complete refreshment of the open flux of the polar caps, that is, straight through-put of open flux from the dayside to the nightside X-lines and convection from the dayside OCB to the nightside OCB.

In the example presented in Figure 7, repeating substorms occurred with a repetition rate of ~30 min. Indeed, these may not be true substorms, but substorm intensifications caused by convection-braking. We speculate that sawtooth events, ~3 hr quasi-periodic intense substorms (Belian et al., 1995), may be an extreme case of braking substorms occurring during strongly driven intervals associated with geomagnetic storms (Walach & Milan, 2015; Walach et al., 2017). In this case, we would place SMCs, repeating substorms, and sawtooth events as a spectrum of responses to periods of prolonged low to high solar wind-magnetosphere coupling; we note that this spectrum of behavior agrees with the ordering of Hubert et al. (2017). On the other hand, isolated substorms are associated with periods when the IMF is sporadically turning northward and southward.

We have not addressed the question of why some substorms commence at high latitudes and others at low latitudes, though Milan, Hutchinson, et al. (2009) suggested that this was associated with the magnetic perturbation produced by the ring current dipolarizing the near-Earth tail. This will be investigated further in a subsequent study.

## References

### Acknowledgments

S.E.M. and J.A.C. were supported by the Science and Technology Facilities Council (STFC), UK, grant no. ST/N000749/1; H.S. was supported by an STFC studentship. M.T.W. was supported by the Natural Environment Research Council (NERC), UK, grant no. NE/P001556/1. The work at the Birkeland Centre for Space Science, University of Bergen, Norway, was supported by the Research Council of Norway/CoE under contract 223252/F50. We thank the AMPERE team and the AMPERE Science Center for providing the Iridium-derived data products; AMPERE products are available online (http://ampere.jhuapl.edu). The OMNI data, including solar wind parameters and geomagnetic indices, were obtained from the GSFC/SPDF OMNIWeb interface (at http://omniweb.gsfc.nasa.gov.) The SuperMAG indices and substorm list was downloaded online (http://supermag.jhuapl.edu). For the ground magnetometer data from which these were derived, we gratefully acknowledge: Intermagnet; USGS, Jeffrey J. Love; CARISMA, PI Ian Mann; CANMOS; The S-RAMP Database, PI K. Yumoto and K. Shiokawa; The SPIDR database; AARI, PI Oleg Troshichev; The MACCS program, PI M. Engebretson, Geomagnetism Unit of the Geological Survey of Canada; GIMA; MEASURE, UCLA IGPP and Florida Institute of Technology; SAMBA, PI Eftyhia Zesta; 210 Chain, PI K. Yumoto; SAMNET, PI Farideh Honary; The institutes who maintain the IMAGE magnetometer array, PI Eija Tanskanen; PENGUIN; AUTUMN, PI Martin Connors; DTU Space, PI Dr. Juergen Matzka; South Pole and McMurdo Magnetometer, PI's Louis J. Lanzarotti and Alan T. Weatherwax; ICESTAR; RAPIDMAG; PENGUIn; British Artarctic Survey; McMac, PI Peter Chi; BGS, PI Susan Macmillan; Pushkov Institute of Terrestrial Magnetism, Ionosphere and Radio Wave Propagation (IZMIRAN); GFZ, PI Juergen Matzka; MFGI, PI B. Heilig; IGFPAS, PI J. Reda; University of L'Aquila, PI M. Vellante; SuperMAG, PI Jesper W. Gjerloev.